\documentclass[conference]{IEEEtran}
\IEEEoverridecommandlockouts

\usepackage{array}
\usepackage[noend]{algpseudocode}
\usepackage{cite}
\usepackage{amsmath,amssymb,amsfonts}
\usepackage{algorithm}
\usepackage{graphicx}
\usepackage{textcomp}

\usepackage[numbers]{natbib}    
\def\BibTeX{{\rm B\kern-.05em{\sc i\kern-.025em b}\kern-.08em
    T\kern-.1667em\lower.7ex\hbox{E}\kern-.125emX}}

\begin{document}
%
\title{Delog: A Privacy Preserving Log Filtering Framework for Online Compute Platforms}

\author{
    \IEEEauthorblockN{
        Amey Agrawal\IEEEauthorrefmark{1},
        Abhishek Dixit\IEEEauthorrefmark{1},
        Namrata A Shettar\IEEEauthorrefmark{1},
        Darshil Kapadia\IEEEauthorrefmark{2}
        \thanks{\IEEEauthorrefmark{2}Work done during an internship at Qubole},\\
        Vikaram Agrawal\IEEEauthorrefmark{1},
        Rajat Gupta\IEEEauthorrefmark{1},
        and Rohit Karlupia\IEEEauthorrefmark{1}
    }
    \IEEEauthorblockA{
        \IEEEauthorrefmark{1}
        Qubole India\\
        Bengaluru, India\\
        Email: ameya@qubole.com
    }

    \IEEEauthorblockA{
        \IEEEauthorrefmark{2}
        BITS Pilani \\
        Vidyavihar, India \\
        Email: f2015113@pilani.bits-pilani.ac.in
    }
}

\maketitle


%
\begin{abstract}

In many software applications, logs serve as the only interface between the application and the developer. However, navigating through the logs of long-running applications is often challenging. Logs from previously successful application runs can be leveraged to automatically identify errors and provide users with only the logs that are relevant to the debugging process. We describe a privacy preserving framework which can be employed by Platform as a Service (PaaS) providers to utilize the user logs generated on the platform while protecting the potentially sensitive logged data. Further, in order to accurately and scalably parse log lines, we present a distributed log parsing algorithm which leverages Locality Sensitive Hashing (LSH). We outperform the state-of-the-art on multiple datasets. We further demonstrate the scalability of Delog on publicly available Thunderbird log dataset with close to 27,000 unique patterns and 211 million lines.

\end{abstract}




\section{Introduction}

Online cloud computing platforms have made large scale distributed computing accessible at affordable prices, leading to a surge in usage of distributed computing frameworks like Apache Spark \cite{spark}, Hive \cite{hive} and Presto \cite{presto}. However, in case of application failures, the user has to navigate through massive amounts of recorded logs to diagnose issues, causing a dip in productivity and unsatisfactory user experience.

In this paper, we describe a log filtering framework, Delog, which identifies anomalous logs at run time in order to minimize the manual effort required by the user in case of failures. We use a simple assumption that any log pattern that frequently occurs in successful application runs is irrelevant to identification of errors. Further, every log line is considered to be a string generated from a template where some tokens are constants while others represent values of certain variables. The process of identifying patterns in log lines is formally known as  log  parsing. Log parsing techniques try to extract unique patterns which correspond to specific system events by recognizing variables.

Some of the existing works in log parsing \cite{logan}, \cite{pop} use distributed computing to solve the problems posed by large volumes of logs generated in production use cases. However, their performance deteriorates with increasing number of unique log patterns. To address this problem, we propose a novel algorithm based on Locality Sensitive Hashing (LSH) which can efficiently handle datasets with large number of patterns. We demonstrate the efficacy of Delog on Hive and Thunderbird \cite{zhu2018tools} datasets where existing techniques perform poorly. We further perform a grid search over the model hyperparameters to exhibit their robustness across datasets.

Most previous studies \cite{lke}, \cite{pop}, \cite{iplom} on log parsing make an assumption that the token count for instances of a given log pattern remains constant. However, we find this assumption to be limiting. Consider the following token sequences, \textit{``ContextHandler Started ServeletContextHandler rdd null AVAILABLE Spark''} and \textit{``ContextHandler Started ServeletContextHandler static Spark''}. Here a single variable (corresponding to `\%s' in the code) is printing multiple tokens and hence, the token count of the two sequences is different though both are instances of a single log pattern, \textit{``ContextHandler Started ServeletContextHandler * Spark''}. We attend to such cases using sequence alignment, which allows us to obtain optimal reduced form of a log pattern and leads to better generalization ability.

Modern applications use multiple third party libraries and frameworks for different problem domains. Therefore, it is impossible to train an algorithm for all possible kinds of log patterns beforehand. A Platform as a Service provider can mitigate this problem by learning patterns from user generated application logs. However, application logs can contain sensitive data including secret keys, usernames and business data. To address this, we propose a privacy preserving framework which uses novel Bloom filter based data encoding technique \cite{bm_lsh} to utilize user logs without compromising user privacy.

To summarize, our key contributions in this work are:

\begin{itemize}
\item We propose a novel log parsing algorithm that can process datasets with large number of patterns while keeping the pattern quality high.
\item We describe a log filtering method which can save time for a user trying to diagnose application issues by navigating through logs.
\item We propose a privacy preserving framework that securely utilizes user generated log data to improve filtering performance.
\item We release \footnote{https://github.com/qubole/qubole-log-datasets} a synthetic dataset which can be used to evaluate the performance of privacy preserving log filtering frameworks in future.
\end{itemize}
The rest of this paper is organized as follows: In Section \ref{sec:log_parsing}, we introduce the log parsing algorithm. Section \ref{sec:log_filtering} presents the log filtering methodology in detail. The privacy preserving framework to learn from user logs is described in section \ref{sec:privacy_preserving}. Evaluation and experimental results of our approach is present in Sections \ref{sec:log_parsing_eval} and \ref{sec:privacy_preserving_eval}. The related works and conclusion are presented in Section \ref{sec:related_works} and Section \ref{sec:conclusion}, respectively.


\section{Log Parsing}
\label{sec:log_parsing}

Log parsing is typically the first step of any log processing system. In order to obtain high quality log patterns, we develop a novel log parsing algorithm which can scale to handle datasets with large number of patterns. Our log parsing technique consists of three stages. The first stage involves preprocessing input log lines to identify trivial variables. These preprocessed lines are then divided into blocks containing similar lines. Finally, each block is reduced to a single log pattern by identifying constant tokens. In the rest of this section, we discuss each step of the parsing algorithm in detail.

\begin{figure}[t]
\centerline{
    \includegraphics[width=9cm]{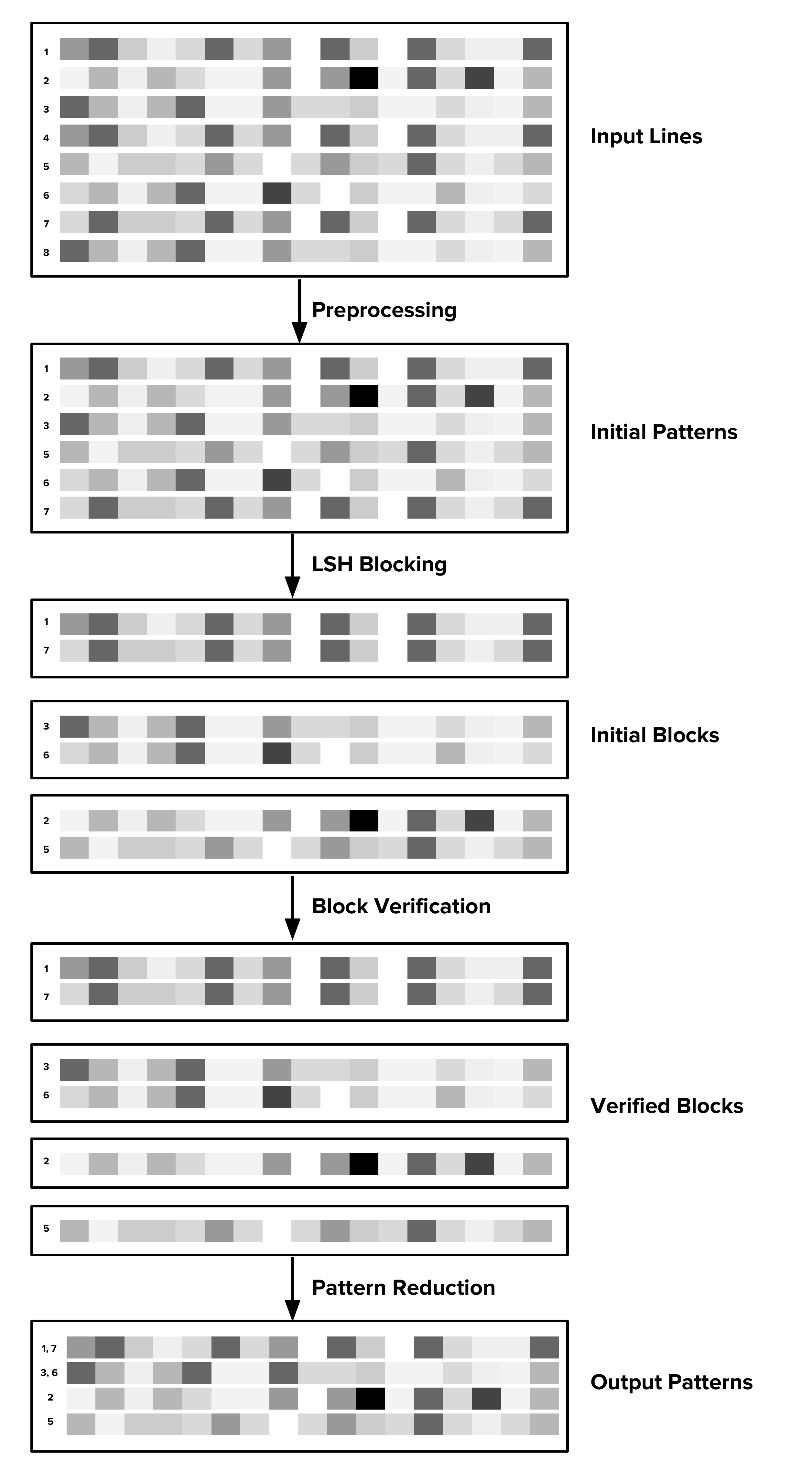}
}
\caption{Log Parsing Pipeline: 
The log lines fed to the pipeline are first preprocessed resulting in the initial set of patterns (Subsection \ref{ssec:preproc}). Preprocessed lines with similar patterns are then put in blocks using minhash-based LSH (Subsection \ref{ssec:lsh_clustering}). These blocks are further verified (Subsection \ref{ssec:lcs_clustering}) using LCS algorithm. The verified blocks are then aligned using a variation of Multiple Sequence Alignment algorithm (Subsection \ref{ssec:seq_align}) and the Alignment Matrix thus formed is reduced to a single pattern based on the strategy described in Subsection \ref{ssec:table_reduce}. Further, the blocks are iteratively reduced by successively repeating the above steps till the number of resulting patterns become constant.}
\label{fig:overview}
\end{figure}

\subsection{Preprocessing of Log Lines}
\label{ssec:preproc}

In the preprocessing stage, we split each input line at the spaces to create a sequence of tokens. First, we filter out easy to identify variables like URLs and file paths from the list of tokens. Then we split the remaining tokens on every non-alphanumeric character and identify numbers, hexadecimal numbers and encoded strings. This helps us to handle tokens like IP Addresses (``127.0.0.1''), dates (``10/01/2018''), and other similar tokens without having to match manually defined regular expressions. All the remaining unidentified tokens are marked as strings. In the next step, consecutive non-string tokens occurring in the pattern are replaced by a wild card. Finally, we remove the duplicate patterns to return a set of unique lines in the processed file. We speed up the preprocessing stage by introducing Least Recently Used (LRU) caches on costly tokenization operations.

\subsection{Blocking Preprocessed Lines}
\label{ssec:lsh_clustering}

We typically observe that the number of unique lines obtained after preprocessing is less than 0.1\% of the original number of lines in the file. However, this number can still be large and finding similar lines by naive comparisons can be costly. To reduce the costly matching operations, we block the preprocessed lines using minhash-based Locality Sensitive Hashing (LSH) \cite{lsh}. Considering the nature of log datasets, we use token-level shingles as opposed to the more traditional character-level shingles. This further helps in reducing the hash compute and comparison time. We create candidate log pattern blocks using the LSH with an empirically chosen Jaccard similarity threshold.

\subsection{Block Verification}
\label{ssec:lcs_clustering}
Since both minhash and LSH are probabilistic data structures, there could be errors in the blocks formed by the method described in \ref{ssec:lsh_clustering}.
We use Longest Common Subsequence (LCS) algorithm to identify any outliers within a block. For a LSH block comprising of \textit{n} patterns $P_1$, $P_{2}$..., $P_{n}$, we regroup them into \textit{k} blocks $B_{1}$, $B_{2}$, ... $B_{j}$ .... $B_{k}$ such that any two patterns in a block $B_{j}$ will satisfy the similarity constraint given by equation \ref{eq:1}. To reduce the number of LCS computations while clustering, we assume a transitive relationship between the patterns within the same LSH block. This implies that if a pattern P satisfies the similarity constraint with any one pattern in the block $B_j$, it will satisfy the constraint for all the patterns in the block $B_j$. Using this transitive property, we iteratively assign each pattern in the original LSH block to a regrouped block which satisfies the constraint. We introduce a new block if we find no suitable block for assignment.
\begin{equation}
\begin{split}
     \label{eq:1}
    \text{LCS}(P, Q).\text{length} - \alpha * \text{max}(P.\text{length}, Q.\text{length}) \ge 0
\end{split}
\end{equation}

where $P$, $Q$ are two patterns and $\alpha$ is an empirically determined constant. 

\subsection{Sequence Alignment}
\label{ssec:seq_align}
By observing the log patterns generated by existing parsing algorithm, we notice that some string variables can consist of more than one token. Methods like LKE \cite{lke} and POP \cite{pop}, check the distribution of tokens at a given position in a block of similar log patterns for reduction. However, such approaches produce sub-optimal results when the sequences do not align. To overcome this limitation, we create a variation of iterative Multiple Sequence Alignment (MSA) algorithm to progressively align sequences. We exploit the shorter lengths and higher similarity between log pattern sequences to simplify and speed up the alignment process. We use Needleman-Wunsch algorithm \cite{needleman} for pairwise alignment of sequences. We sort the patterns in a block by sequence length and then progressively align sequences longest to shortest. At each step, subsets of previously aligned sequences are updated such that the length of all aligned sequences is equal. Details of this step are shown in the Algorithm \ref{alg:msa}.

\begin{algorithm}
\caption{LongestFirstMSA}\label{euclid}
\label{alg:msa}
\begin{algorithmic}[1]
\Procedure{alignSequence}{}

\textbf{Input:} Sequence of patterns \textit{P}

\textbf{Ouput:} Sequence of aligned patterns \textit{A}

\State $\textit{S} \gets \textit{sortByLength(P).reverse}$

\State $\textit{A} \gets \textit{alignPair(S[0], S[1])}$

\For{$i \gets 2$ \textbf{to} \textit{S.length} $-1$}
\State $\textit{alignedPair} \gets \textit{alignPair(A.last, S[i])}$ 

\If {$\textit{alignedPair[0].length} > \textit{A.last.length}$}

\State $\textit{A.removeLast}$
\State $\textit{A.append(alignedPair[0])}$
\State $\textit{A} \gets \textit{alignSequence(A)}$

\EndIf

\State $\textit{A.append(alignedPair[1])}$

\EndFor

\Return $\textit{A}$

\EndProcedure
\end{algorithmic}
\end{algorithm}

\subsection{Alignment Matrix for Pattern Reduction}
\label{ssec:table_reduce} 
\begin{figure}[htbp]
\centerline{
    \includegraphics[width=9cm]{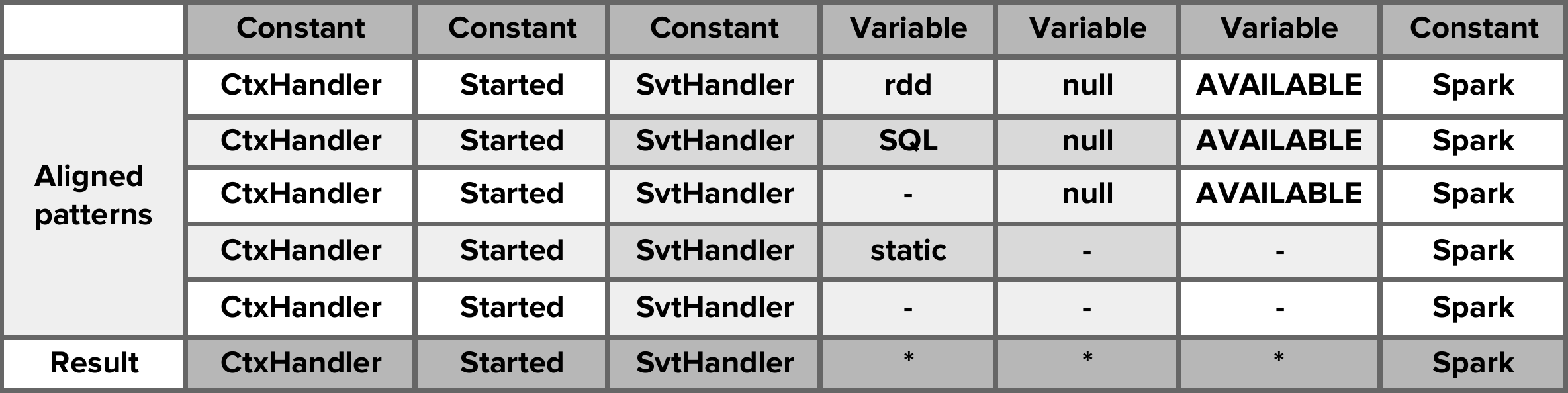}
}
\caption{Pattern Reduction with Alignment Matrix}
\label{fig:table_reduce}
\end{figure}

Consider a block of aligned pattern sequences denoted as $P_0$, $P_1$, ... $P_n$. Each pattern $P_i$ contains $m$ tokens $T_{i1}$, $T_{i2}$, ..., $T_{im}$. We visualize these sequences of patterns as an alignment matrix, where each row represents a pattern in the block as depicted in Figure \ref{fig:table_reduce}. Intuitively, for any given column, if the frequency of a certain token is high, the token position should belong to a constant. On the other hand if a column corresponds to a variable, there should be a lot more variation in the column values. For each column $C_j$ where frequency of mode is denoted as $f_j$, we label it as a constant if it satisfies the heuristic constraint defined in equation \ref{eq:2}.
\begin{equation}
\begin{split}
     \label{eq:2}
    f_j - \beta * n \ge 0
\end{split}
\end{equation}

where \textit{n} is the number of patterns in the matrix and $\beta$ is an empirically determined constant. The remaining columns which do not satisfy equation \ref{eq:2} are labeled as variable columns. 

A row $P_i$ is considered a misfit in the matrix if for any column $C_j$ labeled as a constant, the value of token $T_{ij}$ is not same as the mode of $C_j$. After eliminating misfit rows we simply return the reduced pattern by picking modal value tokens for constant columns and wildcards for variable columns.

\subsection{Iterative reduction}

During the blocking stage, some similar patterns might end up in different blocks due to the approximate nature of LSH. To obtain the optimal set of patterns we iteratively repeat the steps described between subsections \ref{ssec:lsh_clustering} and \ref{ssec:table_reduce} until the number of patterns becomes constant.


\section{Filtering Error Logs}
\label{sec:log_filtering}

One of the key applications of log parsing techniques has been to filter out the anomalies from the logs of failed application runs. This helps in easier and swift error detection and in turn can save a lot of time for developers. This is especially true for applications like Apache Spark and Apache Hive where the size of logs can be as large as tens of gigabytes, making it nearly impossible to manually identify the errors from these logs. We identify the commonly occurring patterns by training Delog on logs of successful application runs. Then, we can filter the lines matching these patterns from the log files and in turn we will be left with anomalies, error messages and stack traces. The subsequent subsections describe the anomaly detection algorithm in further detail.

\subsection{Learning Patterns}
\label{ssec:pattern_selection}

To initialize the filtering model for a given application, we parse logs of previously successful runs of the application. However many a time, logs of successful job runs also contain some intermittent failures which succeed on retries. Hence, to avoid missing some key anomalies and errors in the the log file, not all of the patterns learned are used for filtering. We use a greedy approach where we first sort the patterns in our training data set on the basis of frequency and then we iteratively pick the patterns till we achieve a ninety eight percent coverage of the training dataset. The number of patterns selected through this approach will vary for each application type. For example, we observe that for the Spark dataset, we can achieve ninety eight percent coverage through only top five percent patterns. Moreover, we also observe that certain log lines like application start and application termination appear only once in each file but are present in all the files. Therefore, we make sure to include those patterns as well which are present in seventy percent log files or more.

\subsection{Preprocessing of Input Lines}
\label{ssec:preproc_filter}

The first step of the filtering algorithm is same as the first step of the training algorithm where we identify meaningful tokens from the log lines and convert the log lines into patterns. We cache these preprocessed patterns in order to avoid reprocessing the identical log lines in the subsequent stages. 

\subsection{Candidate Pair Identification}
\label{ssec:lsh_filter}

In this step, we use minhash LSH \cite{lsh} to match the lines obtained from the preprocessing stage with the set of patterns obtained from the training stage of the algorithm. For each preprocessed line, we identify candidate pairs by querying LSH for an empirically chosen Jaccard similarity threshold.

\subsection{LCS Matching}
\label{ssec:lcs_filter}

To concretely identify a valid match, we scan the candidate pairs to find the first pattern which satisfies the constraint given by equation \ref{eq:1}. If the constraint is satisfied by at least one candidate pattern, we conclude that the corresponding log line is not of much relevance for error detection. Otherwise, we infer that the log line is an anomaly and should be included in the filtered logs.

\subsection{Frequency Based Filtering}
\label{ssec:feq_fil}

The list of patterns generated from the training stage of the algorithm may not include all the possible log patterns which can be generated from a given application type. This is because the generated logs can depend on a variety of factors like application configurations, environment variables, user behavior etc. Therefore, it is possible to encounter log lines which do not match our trained data set but are of little value for anomaly and error detection. In order to avoid this, we keep a frequency filter for every  unmatched log line $L_i$ such that,
\begin{equation}
\begin{split}
    \label{eq:3}
    f_i - \gamma \le 0
\end{split}
\end{equation}
Where $f_i$ is the frequency of the log line $L_i$ and $\gamma$ is a constant. We filter out the log lines that do not satisfy the constraint defined in equation \ref{eq:3} from our output and treat them similar to a pattern from the training set. The value of the constant $\gamma$ needs to be determined empirically and depends on the application type. For example, we observe that two hundred and fifty is a suitable value for Spark applications.


\section{Privacy Preserving Framework}
\label{sec:privacy_preserving}

In subsection \ref{ssec:feq_fil} we present a way to improve performance on user logs in wild by tracking the frequency of preprocessed patterns. However, such an approach cannot re-utilize knowledge from previous application runs to improve the results in future. In this section, we propose a framework which can learn from user generated logs while preserving user privacy at the same time.

\begin{figure}[htbp]
\centerline{
    \includegraphics[width=7.5cm]{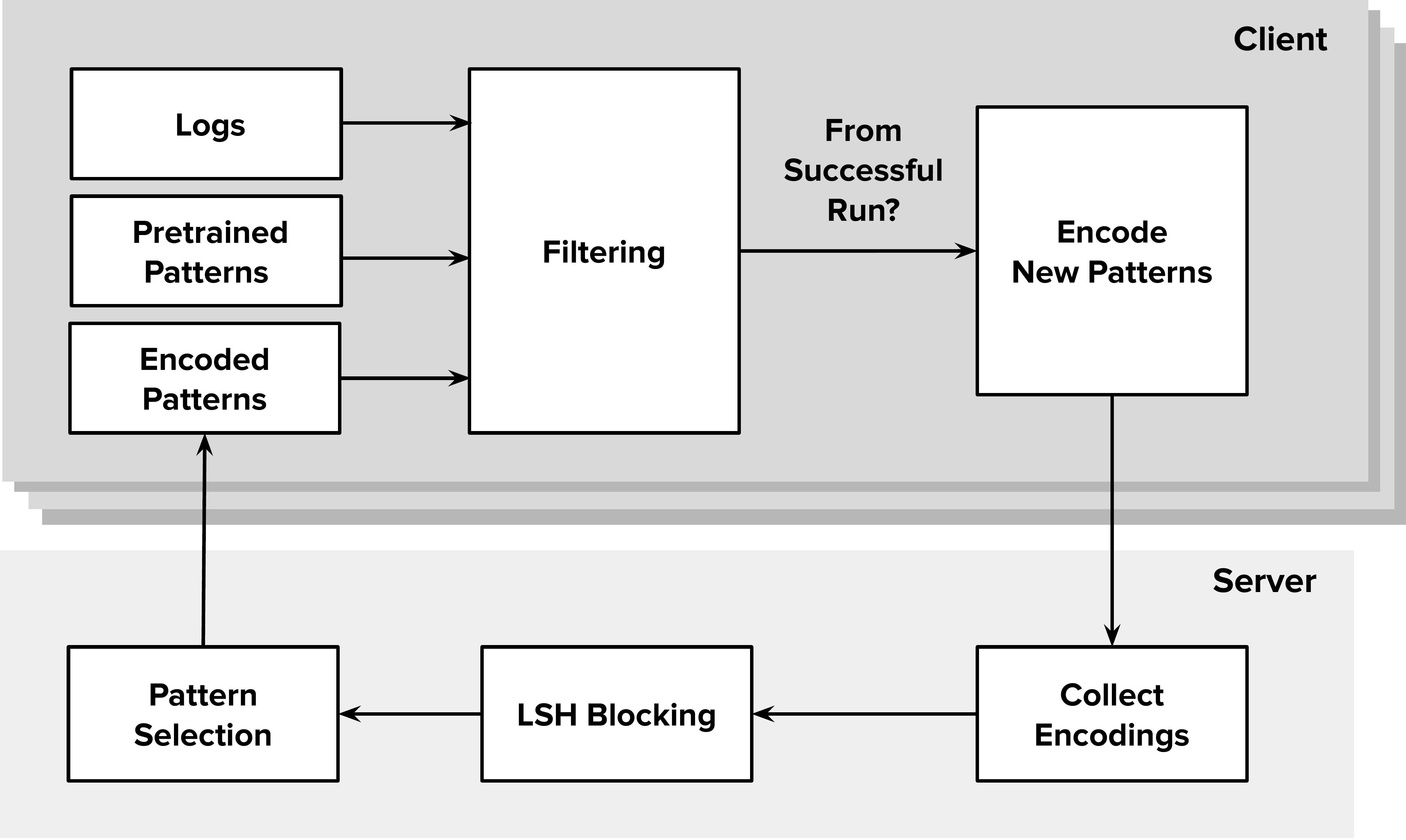}
}
\caption{Privacy Preserving Learning in Client-Server Architecture: The platform provider aggregates and encodes patterns from successful application runs with Bloom filter bitmaps on the client side, which are then sent to a central repository (Subsection \ref{ssec:pp_setup}). After collecting  pattern encodings from clients, server creates a block of similar patterns using LSH (Subsection \ref{ssec:pp_server}). Further, blocks are retained based on their total frequency (Subsection \ref{ssec:pattern_selection}). The clients can now use these encoded log patterns along with pretrained patterns to filter through their logs (Subsection \ref{ssec:pp_filtering}). Any new pattern is encoded at the client and goes through the same series of steps. These resulting patterns which consist of the anomalies, errors and stacktrace are made available for analysis at the client side. }
\label{fig:client_server}
\end{figure}

\subsection{Client-Server Setup}
\label{ssec:pp_setup}

On an online cloud computing platform, hundreds of users run their applications at any given point of time. The platform provider can run a service on the user systems which would aggregate and encode patterns from logs of successful application runs in a homomorphic fashion. These encoded patterns are then sent to a central repository. During the future runs of the application, the encoded log patterns can be fetched from the repository and utilized for matching.

\subsection{Encoding of Log Patterns}
\label{ssec:encode_patterns}
\begin{figure}[htbp]
\centerline{
    \includegraphics[width=9cm]{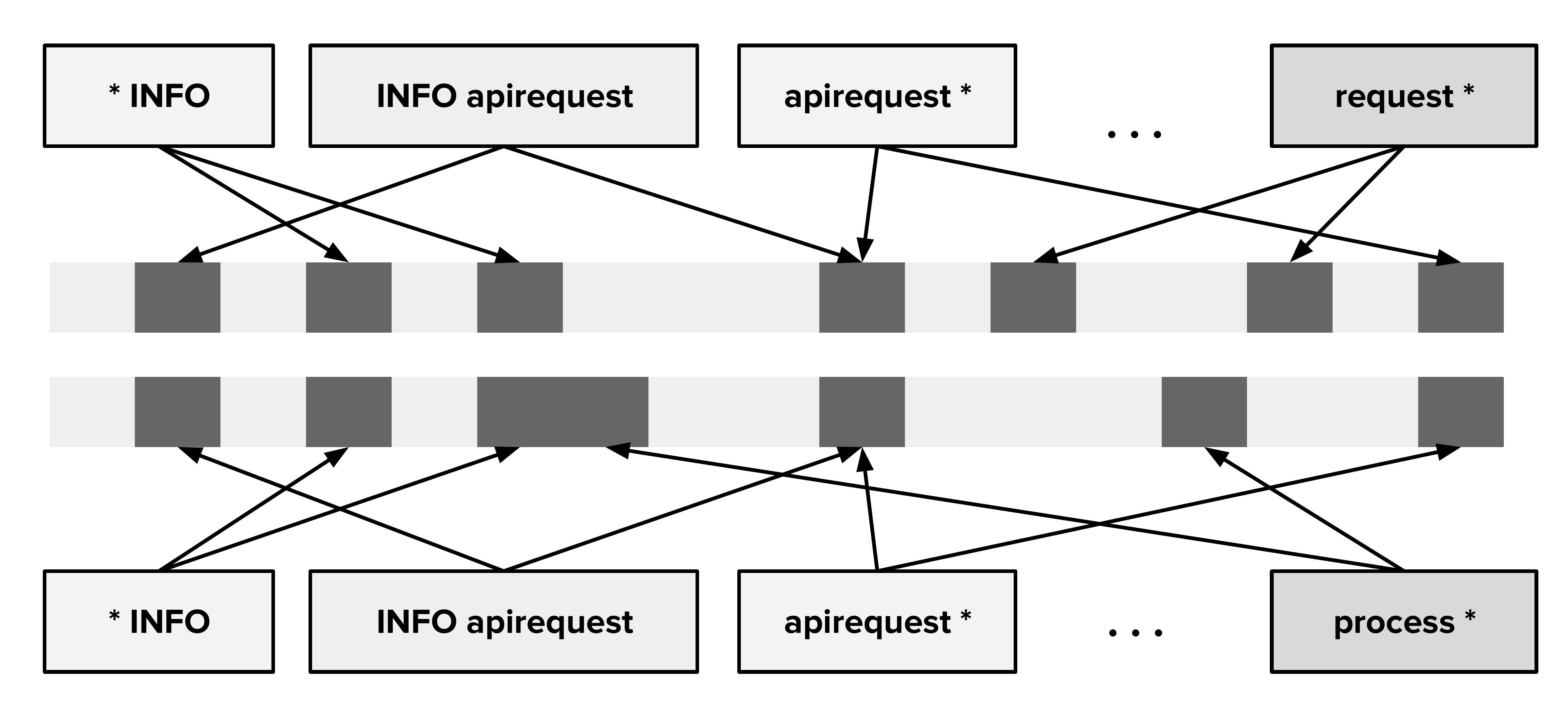}
}
\caption{Privacy preserving encoding of log patterns with Bloom filter bitmaps: The figure shows comparison between bloom filter bitmaps of two log patterns. Each token level shingle of a pattern is mapped to two positions in the bloom filter bitmap using the two hash functions (depicted by the arrows). The patterns are then compared by computing the Jaccard similarity between the two bitmaps.}
\label{fig:bloom_filter}
\end{figure}

Schnell et al. \cite{bm_distance_preserve} in their work on record linking, show a novel use of bloom filters for homomorphic encoding of strings. For each pattern we create a bitmap by inserting hashes of token level shingles (n-grams) into a bloom filter as shown in Figure \ref{fig:bloom_filter}. It is extremely difficult to reconstruct input pattern from such an encoding. To compare any two patterns we can directly compute the Jaccard similarity of their corresponding bitmaps.

\subsection{Processing at Server}
\label{ssec:pp_server}
After collecting the pattern encodings from all clients, the server creates blocks of similar patterns using LSH. Total frequency of each block is computed and the patterns are selected using a method similar to the one described in subsection \ref{ssec:pattern_selection}. These selected patterns are then communicated back to the clients. The LSH is configured with a high Jaccard similarity threshold, so as to minimize false positives.

\subsection{Inference Procedure}
\label{ssec:pp_filtering}
In addition to the method described in section \ref{sec:log_filtering}, the clients use encoded log patterns received from the server. A log line which does not match to any of the pretrained patterns is then encoded and looked for in the set of encoded patterns. A LSH is used to speed up the search for candidate pairs and we validate the candidate patterns by computing Jaccard similarity between the bitmaps. Any pattern with a valid match is then excluded from the output.

\begin{figure}[htbp]
\centerline{
    \includegraphics[width=6cm]{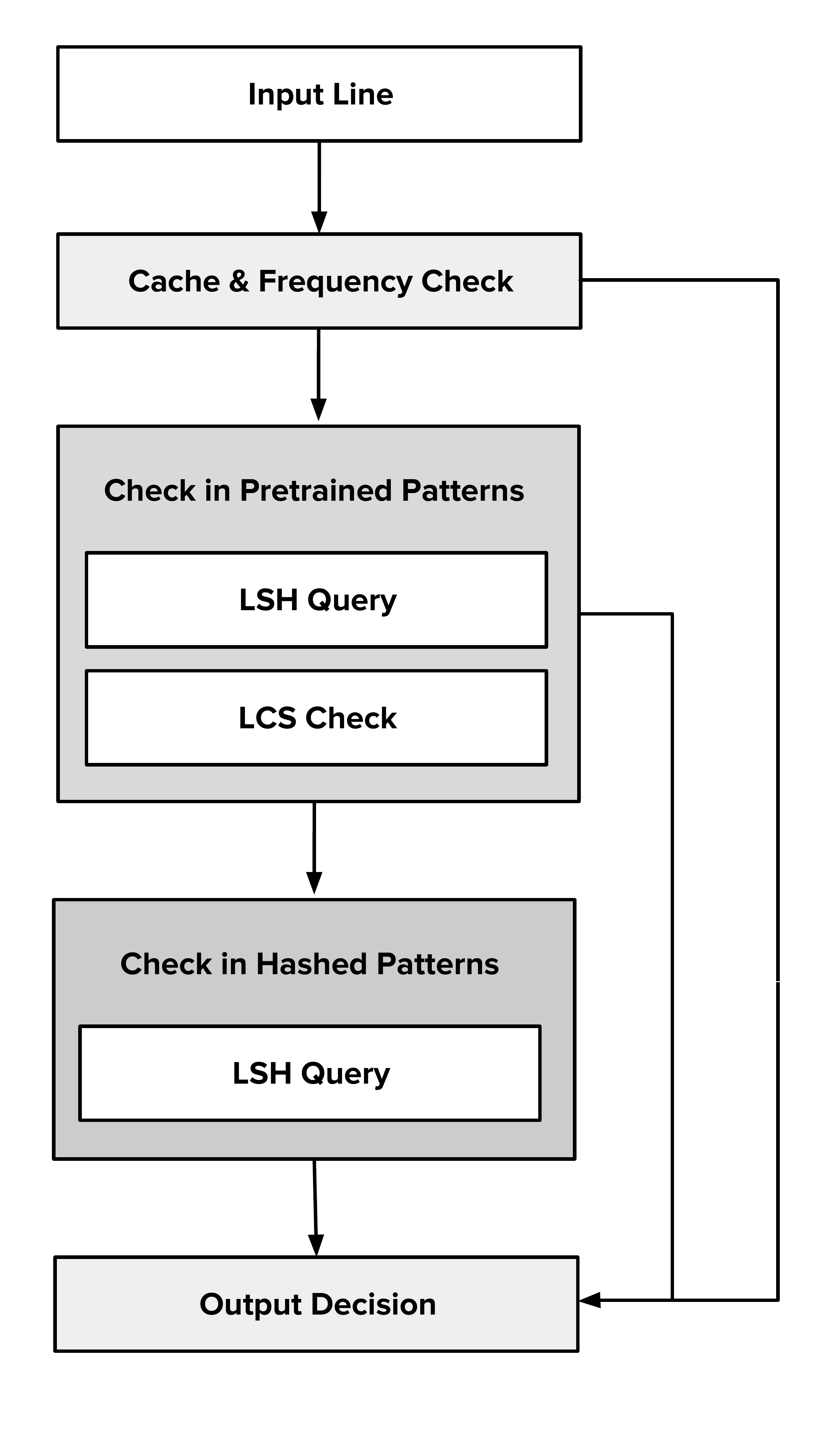}
}
\caption{Log Filtering Pipeline:
We first identify meaningful tokens from the log lines and convert them into patterns (Subsection \ref{ssec:preproc_filter}). During this step, we cache preprocessed patterns to avoid reprocessing of identical lines. In the next step, we match the preprocessed lines with the patterns identified from training stage first using minhash based LSH (Subsection \ref{ssec:lsh_filter}) and then using LCS filtering (Subsection \ref{ssec:lcs_filter}). We further match the log lines which do not match any pretrained patterns with the encoded patterns received from the privacy preserving server. (Subsection \ref{ssec:pp_filtering}). Certain unmatched log lines may still have little value for anomaly detection. So in the end, we keep a frequency filter which further filters out the log lines satisfying a given frequency threshold (Subsection \ref{ssec:feq_fil}). The remaining unmatched log lines are considered as anomalies and printed in the output.}
\label{fig:inference_pipeline}
\end{figure}


\section{Evaluating Log Parser}
\label{sec:log_parsing_eval}

\subsection{Experimental Setup}

In this section, we compare Delog with some notable log parsing algorithms like SHISO \cite{shiso}, Spell \cite{spell}, Drain \cite{drain}, IPLoM \cite{iplom} and Logan \cite{logan}. We run our experiments on Amazon EC2 r3.8xlarge instances with 32 virtual cores. We run Delog as a Spark application deployed in Yarn client mode on a five node cluster with 128 cores to take advantage of the distributed processing.

We omit results for some algorithms on specific datasets if they fail to finish processing in a reasonable time.

\subsection{Datasets}
We evaluate Delog on 10 real-world datasets including seven publicly available datasets and three proprietary datasets. Six of these datasets comprise of distributed system logs, namely: Zookeeper \cite{drain}, HDFS \cite{jordanji}, HDFS2 \cite{zhu2018tools}, Spark, Presto and Hive. BGL \cite{BGL}, Thunderbird and HPC \cite{hpc} datasets comprise of system logs from super-computers. Finally, Windows \cite{zhu2018tools} dataset is made of operating system logs generated from computers running Windows 7. Each of Spark, Presto and Hive datasets are generated by thousands of unique queries run on Amazon Elastic Compute Cloud clusters used for analytics inside our organization. Since the Spark, Hive and Presto datasets are business sensitive and closed, our experiments on publicly available datasets of Thunderbird, Windows and HDFS made available by \cite{zhu2018tools} provide scope for easier comparison with other distributed techniques in future.

The Thunderbird logs are largest by volume with a total of two hundred and eleven millions lines. Further more, it contains close to twenty seven thousand patterns due of which the existing distributed log parsing algorithms fail to parse in reasonable time. Our experiments with the Thunderbird dataset are the first reported results of log parsing on a dataset of such scale. Table \ref{table:datasets} contains the details of size and number of log lines in these datasets.

\begin{table}[htbp]
\caption{Summary of Datasets}
\begin{center}
\bgroup
\def\arraystretch{1.5}
\begin{tabular}{|c|c|c|}
\hline
\textbf{\textit{Dataset}} & \textbf{\textit{Number of Lines}}& \textbf{\textit{Total Size}} \\
\hline
\textbf{Zookeeper} & 74,380 & 10 MB \\
\hline
\textbf{HPC} & 433,490 & 32 MB\\
\hline
\textbf{BGL} & 4,747,963 & 109 MB\\
\hline
\textbf{Presto-5.2M} & 5,200,000 & 1.48 GB\\
\hline
\textbf{HDFS} & 11,175,629 & 1.57 GB \\
\hline
\textbf{Spark-13M} & 13,000,000 & 2.10 GB\\
\hline
\textbf{Presto} & 16,222,793 & 4.73 GB\\
\hline
\textbf{Hive} & 39,010,740 & 7.9 GB \\
\hline
\textbf{HDFS2} & 71,118,073 & 16.06GB \\
\hline
\textbf{Spark} & 96,503,051 & 14.75 GB\\
\hline
\textbf{Windows} & 114,608,388 & 26.09GB \\
\hline
\textbf{Thunderbird} & 211,212,192 & 29.60 GB \\
\hline

\hline
\end{tabular}
\egroup
\end{center}
\label{table:datasets}
\end{table}

\subsection{Evaluation Metric}

Earlier works quantify the accuracy of log parser using metrics like F-measure. However, this involves manually obtaining patterns for the dataset. Since this is not feasible for large datasets like Hive \cite{hive} and Spark \cite{spark}, we decide to use the quality loss metric used in Logan \cite{logan}. Quality loss is computed using equation \ref{eq:6} and it penalizes those patterns in which meaningful tokens are lost and converted into wildcards. Logan \cite{logan} uses another function called length factor which penalizes the algorithms for generating too many patterns. A naive parser which returns every single line in the dataset without any processing would have a zero quality loss, but would receive high penalty through length factor. On the other hand, an ideal log parser would minimize the number of log patterns while keeping the quality loss low. Although Logan combines quality loss and length factor to come up with a new loss function, we observe that it becomes difficult to capture the impact of quality factor this way. Therefore, we separately compare algorithms for their quality scores and the number of patterns they generate.

\begin{equation}
\label{eq:5}
\text{averageTokensLost}(p) = \text{averageMatchLength}(p) - p.\text{length}
\end{equation}
\begin{equation}
\label{eq:6}
    \text{qualityLoss}(P) = \frac{1}{P.{\text{length}}} \sum_{i = 0}^{P.{\text{length}}} (\frac{\text{averageTokensLost}(p_i)}{p_i.{\text{length}}}) ^ 2
\end{equation}
\\
Where, where $P$ is the set of all identified patterns and $\text{averageMatchLength}(p)$ is the average length of all sequences matched to pattern $p$ during training.

\subsection{Observations}

\subsubsection{Parse Quality}

Looking at the tables \ref{table:loss} and \ref{table:num_patts}, we observe that Delog consistently produces high quality patterns across datasets with a reasonable number of output patterns. High loss value along with fewer patterns in SHISO \cite{shiso} suggests that many meaningful tokens are lost. IPLoM \cite{iplom} performs well for HDFS \cite{hdfs} dataset but it fails in terms of quality of patterns when it comes to other datasets. Drain's algorithm over simplifies log patterns leading to steep decline in the parse quality as we can see in the Presto dataset.

\begin{table*}[bt]
\begin{center}
\caption{Loss measurement for Delog and other methods}
\bgroup
\def\arraystretch{1.5}
\begin{tabular}{|>{\centering\arraybackslash}p{1cm}|>{\centering\arraybackslash}p{1cm}|>{\centering\arraybackslash}p{1cm}|>{\centering\arraybackslash}p{1cm}|>{\centering\arraybackslash}p{1cm}|>{\centering\arraybackslash}p{1cm}|>{\centering\arraybackslash}p{1cm}|>{\centering\arraybackslash}p{1cm}|>{\centering\arraybackslash}p{1cm}|>{\centering\arraybackslash}p{1cm}|>{\centering\arraybackslash}p{1cm}|>{\centering\arraybackslash}p{1cm}|>{\centering\arraybackslash}p{1cm}|}
\hline
& \textbf{\textit{Zook-eeper}} & \textbf{\textit{HPC}} & \textbf{\textit{BGL}} & 
\textbf{\textit{Presto-5.2M}} & \textbf{\textit{HDFS}} & \textbf{\textit{Spark-13M}} 
& \textbf{\textit{Presto}} & \textbf{\textit{Spark}} &
\textbf{\textit{Hive}} &
\textbf{\textit{Thund-erbird}} &
\textbf{\textit{Wind-ows}} &
\textbf{\textit{HDFS2}}  \\
\hline
\textbf{SHISO} & 2.7385 & 4.6333 & - & - & - & - & - & - & - & - & - & - \\
\hline
\textbf{Spell} & \textbf{0.0002} & 0.0091 & - & - & - & - & - & - & - & - & - & - \\
\hline
\textbf{Drain} & 0.0042 & 0.1 & 0.0849 & 120.9164 & 0.0013 & 0.6445 & - & - & - & - & - & -\\
\hline
\textbf{IPLoM} & 0.0053 & 1.8697 & 0.0091 & 7.3184 & \textbf{0.0004} & 7.13 & - & - & - & - & - & - \\
\hline
\textbf{Logan} & 0.0377 & 0.0598 & 0.0211 & \textbf{0.0123} & 0.0107 & 0.0056 & \textbf{0.0069} & 0.0048 & \textbf{0.0135} & - & 0.0437 & 0.0057 \\
\hline
\textbf{Delog} & \textbf{0.0002} & \textbf{0.0005} & \textbf{0.0019} & 0.0144 & 0.0009 & \textbf{0.0025} & 0.0104 & \textbf{0.0038} & 0.0159 & \textbf{0.0012} & \textbf{0.0397} & \textbf{0.0013}\\
\hline

\end{tabular}
\label{table:loss}
\egroup
\end{center}

\end{table*}

\begin{table*}[bt]
\begin{center}
\caption{Number of patterns identified by Delog and other methods}
\bgroup
\def\arraystretch{1.5}
\begin{tabular}{|>{\centering\arraybackslash}p{1cm}|>{\centering\arraybackslash}p{1cm}|>{\centering\arraybackslash}p{1cm}|>{\centering\arraybackslash}p{1cm}|>{\centering\arraybackslash}p{1cm}|>{\centering\arraybackslash}p{1cm}|>{\centering\arraybackslash}p{1cm}|>{\centering\arraybackslash}p{1cm}|>{\centering\arraybackslash}p{1cm}|>{\centering\arraybackslash}p{1cm}|>{\centering\arraybackslash}p{1cm}|>{\centering\arraybackslash}p{1cm}|>{\centering\arraybackslash}p{1cm}|}
\hline
& \textbf{\textit{Zook-eeper}} & \textbf{\textit{HPC}} & \textbf{\textit{BGL}} & 
\textbf{\textit{Presto-5.2M}} & \textbf{\textit{HDFS}} & \textbf{\textit{Spark-13M}} & \textbf{\textit{Presto}} & \textbf{\textit{Spark}} & \textbf{\textit{Hive}} & \textbf{\textit{Thund-erbird}} & \textbf{\textit{Wind-ows}} & \textbf{\textit{HDFS2}} \\
\hline
\textbf{SHISO} & 42 & 90 & - & - & - & - & - & - & - & - & - & - \\
\hline
\textbf{Spell} & 168 & 230 & - & - & - & - & - & - & - & - & - & - \\
\hline
\textbf{Drain} & 89 & 147 & 1202 & 1466 & 48 & 1001 & - & - & - & - & - & -\\
\hline
\textbf{IPLoM} & 91 & 119 & 2765 & 3748 & 41 & 738 & - & - & - & - & - & - \\
\hline
\textbf{Logan} & 100 & 138 & 261 & 2391 & 48 & 895 & 5514 & 2979 & 6432 & - & 2595 & 180 \\
\hline
\textbf{Delog} & 117 & 306 & 1896 & 2829 & 45 & 808 & 461 & 852 & 7887 & 27784 & 3415 & 163\\
\hline
\end{tabular}
\egroup
\label{table:num_patts}
\end{center}
\end{table*}

\subsubsection{Run Time}

Table \ref{table:runtime} shows that for most of the datasets, Delog outperforms other algorithms in run time efficiency. Since Delog uses minhash LSH to identify pattern blocks, we see that it performs better than the previous state-of-the-art, Logan\footnote{We obtain additional performance improvements over results reported by authors \cite{logan} by optimizing the number of shards such that parallelism is improved.} \cite{logan} especially for large datasets. For the Hive dataset which has high number of patterns, Delog is \textbf{\textit{5x}} than Logan. Further, we find that Delog is the only log algorithm which can process the Thunderbird dataset in respectable amount of time because of its sub-linear complexity in number of patterns.

\begin{figure}[htbp]
\centerline{
    \includegraphics[width=7.5cm]{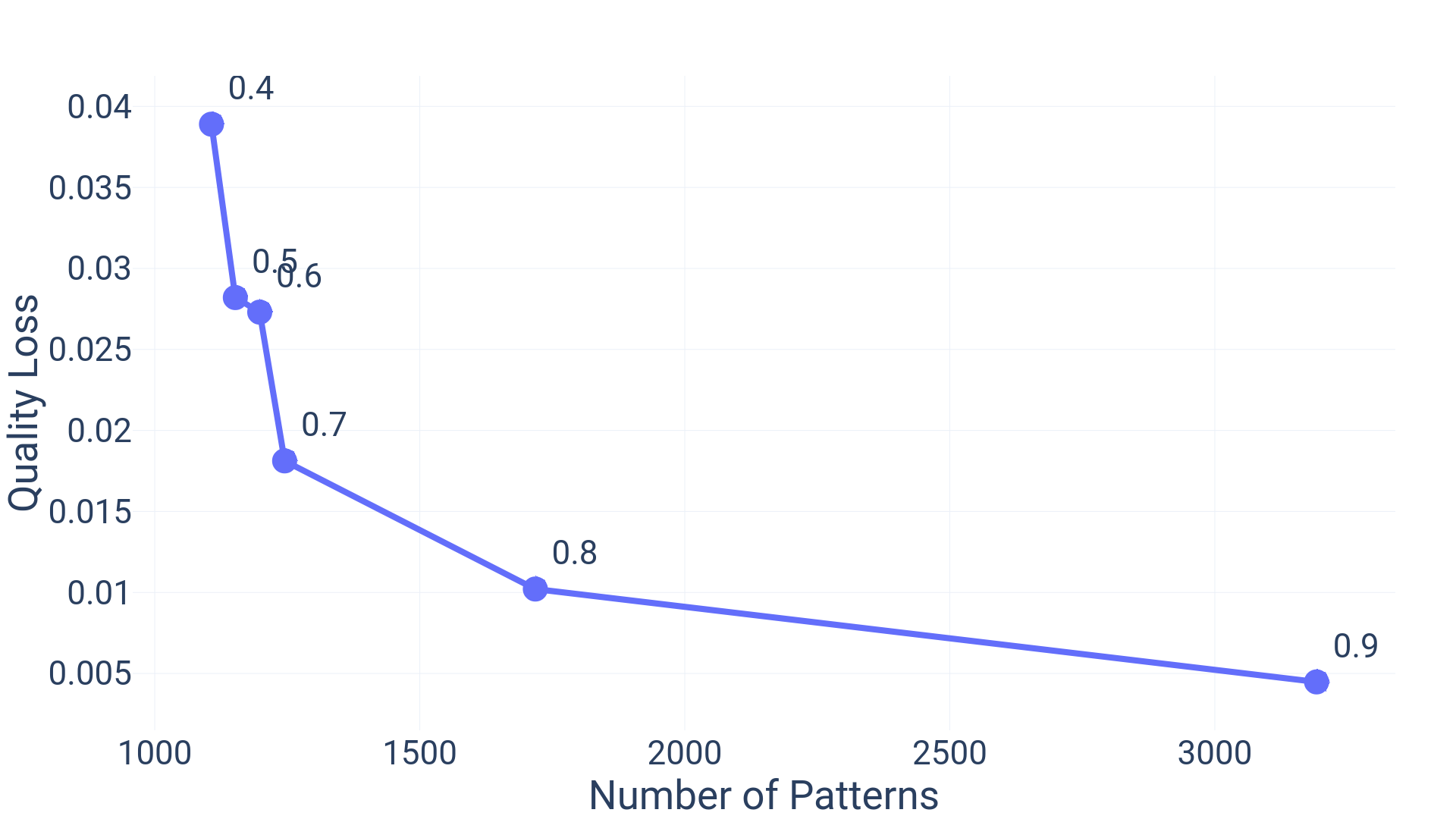}
}
\caption{Variation of number of patterns and quality loss with change in Jaccard similarity threshold during training.}.
\label{fig:jaccard}
\end{figure}

\subsubsection{Impact of Preprocessing}

Table \ref{table:stages} shows number of initial patterns obtained after the preprocessing stage. In most datasets, we see that the number of patterns after preprocessing is less than 0.5\% of the original lines in the dataset. This can be attributed to the large number of numerical and URL variables in the log patterns. The rest of our training pipeline is dedicated to identifying string variables. Depending on the nature of the dataset, we notice a two to three fold reduction in the number of patterns after the iterative reduction step.

\begin{table}
\begin{center}
\caption{Number of Patterns After Each Stage}
\bgroup
\def\arraystretch{1.5}
\begin{tabular}{|c|>{\centering\arraybackslash}p{1.5cm}| >{\centering\arraybackslash}p{1.3cm}|>{\centering\arraybackslash}p{1.5cm}|}
\hline
\textbf{\textit{Dataset}} & \textbf{\textit{Original lines}} & \textbf{\textit{After Preprocessing}} & \textbf{\textit{Output Patterns}} \\
\hline
\textbf{BGL} & 4747963 & 3748 (0.07\%) & 1896 (50.58\%) \\
\hline
\textbf{HPC} & 433490 & 947 (0.21\%) & 306 (32.31\%) \\
\hline
\textbf{Zookeeper} & 74380 & 122 (0.16\%) & 117 (95.90\%)  \\
\hline
\textbf{HDFS} & 11175629 & 47 & 45 (95.74\%) \\
\hline
\textbf{Spark-13m} & 13000000 & 1187 & 808 (68.07\%) \\
\hline
\textbf{Presto-5.2m} & 5200000 & 6566 (0.12\%) & 2829 (43.08\%) \\
\hline
\end{tabular}
\label{table:stages}
\egroup
\end{center}
\end{table}

\subsubsection{Tuning MinHash LSH}
\label{ssec:lsh_tuning}

We use Presto-5.2m dataset to demonstrate parameter tuning for pattern selection using quality loss and number of patterns. Figure \ref{fig:jaccard} shows the impact of changing Jaccard Similarity threshold on the number of output patterns and the quality loss. We see that using a Jaccard similarity threshold of 0.7 for LSH blocking gives an optimal number of patterns along with a desirable quality loss value. Further increasing the Jaccard similarity threshold value leads to a steep increase in number of patterns for a marginal improvement in the quality loss. Similarly, if we decrease the Jaccard similarity threshold in order to reduce the number of patterns, we see that the quality of our patterns significantly goes down.

\subsubsection{Tuning LCS Matching Fraction}
\label{ssec:lcs_tuning}

\begin{figure}[htbp]
\centerline{
    \includegraphics[width=7.5cm]{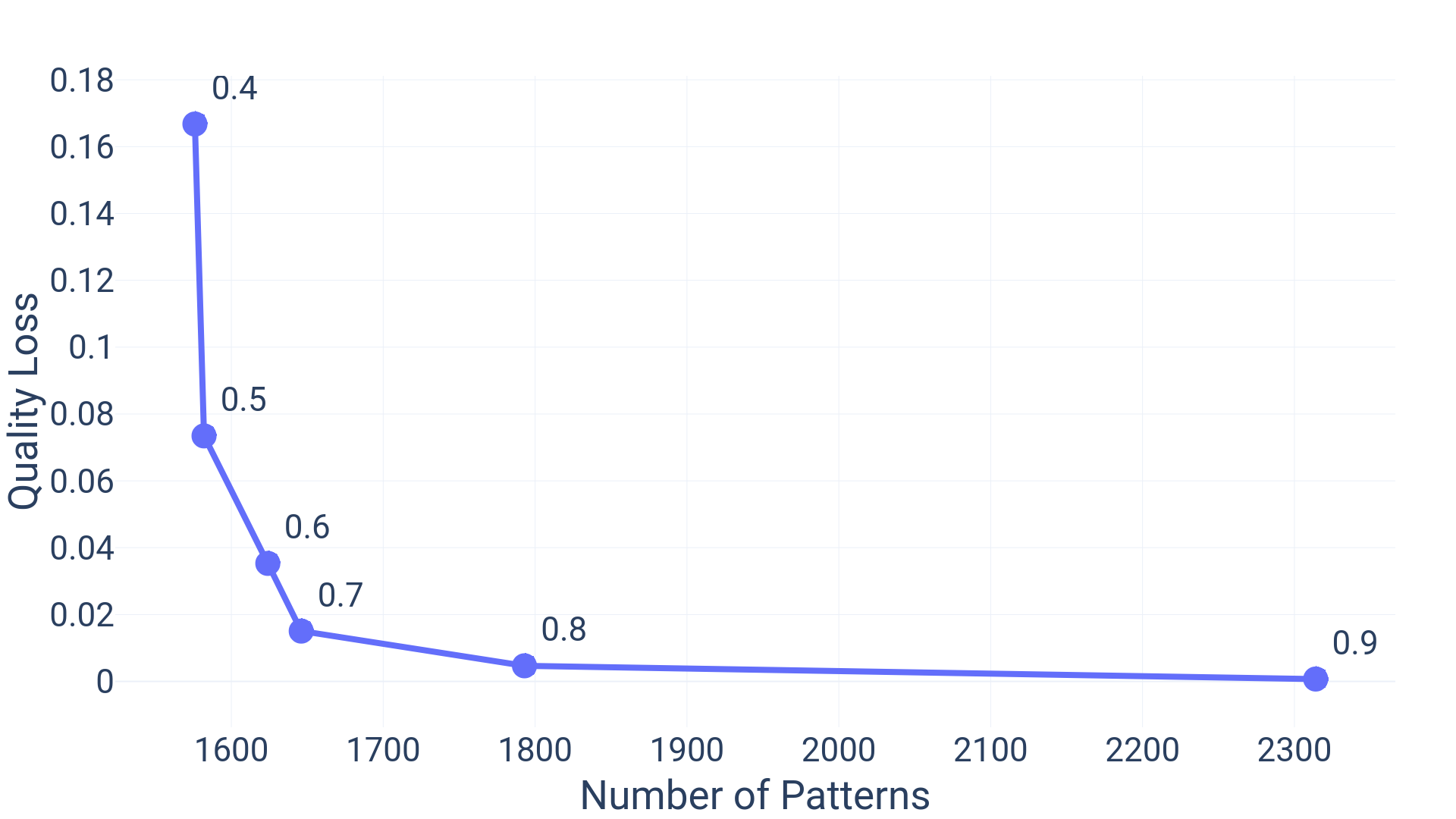}
}
\caption{Variation in number of patterns and quality loss with change in LCS matching threshold ($\alpha$) during training.}
\label{fig:lcs}
\end{figure}

The LCS based regrouping of patterns described in subsection \ref{ssec:lcs_clustering} helps in identification of incorrectly blocked patterns. We observe from figure \ref{fig:lcs} that changing LCS matching threshold ($\alpha$) from 0.4 to 0.7 leads to a significant reduction in quality loss. However, further increase in LCS matching fraction from 0.8 to 0.9 leads to an enormous increase in number of patterns without any substantial dip in the quality loss. Though the optimal value of LCS matching fraction will vary from one dataset to another and needs to be determined empirically.

\begin{table*}[tb]
\begin{center}
\caption{Running time performance of Delog and other methods (in Seconds)}
\bgroup
\def\arraystretch{1.5}
\begin{tabular}{|>{\centering\arraybackslash}p{1cm}|>{\centering\arraybackslash}p{1cm}|>{\centering\arraybackslash}p{1cm}|>{\centering\arraybackslash}p{1cm}|>{\centering\arraybackslash}p{1cm}|>{\centering\arraybackslash}p{1cm}|>{\centering\arraybackslash}p{1cm}|>{\centering\arraybackslash}p{1.2cm}|>{\centering\arraybackslash}p{1cm}|>{\centering\arraybackslash}p{1.1cm}|>{\centering\arraybackslash}p{1cm}|>{\centering\arraybackslash}p{1cm}|>{\centering\arraybackslash}p{1cm}|}
\hline
& \textbf{\textit{Zook-eeper}} & \textbf{\textit{HPC}} & \textbf{\textit{BGL}} & 
\textbf{\textit{Presto-5.2M}} & \textbf{\textit{HDFS}} & \textbf{\textit{Spark-13M}} 
& \textbf{\textit{Presto}} & \textbf{\textit{Spark}} &
\textbf{\textit{Hive}} & \textbf{\textit{Thund-erbird}} & \textbf{\textit{Wind-ows}} & \textbf{\textit{HDFS2}}\\
\hline
\textbf{SHISO} & 129.69 & 656.90 & - & - & - & - & - & - & - & - & - & - \\
\hline
\textbf{Spell} & 13.60 & 76.69 & - & - & - & - & - & - & - & - & - & - \\
\hline
\textbf{Drain} & 3.48 & 16.26 & 156.17 & 436.07 & 559.67 & 7044.18 & - & - & - & - & - & - \\
\hline
\textbf{IPLoM} & 2.18 & 11.25 & 93.23 & 583.55 & 379.03 & 896.22 & - & - & - & - & - & - \\
\hline
\textbf{Logan} & \textbf{1.13} & \textbf{3.78} & 24.16 & 68.34 & 53.39 & 119.03 & 21.38 & \textbf{42.76} & 1660.54 & - & 179.87 & \textbf{29.03} \\
\hline
\textbf{Delog} & 2.03 & 4.23 & \textbf{16.08} & \textbf{27.27} & \textbf{27.42} & \textbf{48.59} & \textbf{21.13} & 84.16 & \textbf{313.96} & \textbf{225.71} & \textbf{109.54} & 41.27\\
\hline
\end{tabular}
\label{table:runtime}
\egroup
\end{center}
\end{table*}

\subsection{Robustness of Hyperparameters}
Our model requires tuning a number of hyperparameters such as LSH Jaccard Similarity threshold, LCS matching fraction and LSH Number of permutations. While working with massive log datasets, tuning hyperparameters could be costly and time consuming. Hence, we use a fixed set of hyperparameters as described in Table \ref{table:param} for all the reported results. We also perform an exhaustive grid search over Zookeeper, HPC, BGL and HDFS datasets to identify optimal hyperparameters by finding the elbow point. Table \ref{table:opti_sol} shows that the optimal hyperparameters across the evaluated datasets have similar values, thus demonstrating the robustness of these hyperparameters.

\begin{table}[htbp]
\caption{Summary of Evaluation Parameters}
\begin{center}
\bgroup
\def\arraystretch{1.5}
\begin{tabular}{|>{\centering\arraybackslash}m{2.5cm}|c|}
\hline
\textbf{\textit{Parameter}} & \textbf{\textit{Value}} \\
\hline
\textbf{$\alpha$} & 0.65 \\
\hline
\textbf{LSH Shingles} & 2-gram \\
\hline
\textbf{LSH Number of Permutations} & 100 \\
\hline
\textbf{LSH Jaccard Threshold} & 0.75 \\
\hline
\textbf{$\beta$} & 0.7 \\
\hline
\textbf{$\gamma$} & 0.7 \\
\hline

\hline
\end{tabular}
\egroup
\end{center}
\label{table:param}
\end{table}

\begin{table}[htbp]
\caption{Range of Parameters Used for Grid Search}
\begin{center}
\bgroup
\def\arraystretch{1.5}
\begin{tabular}{|>{\centering\arraybackslash}m{2.5cm}|c|c|c|}
\hline

\textbf{\textit{Parameter}}
& \textbf{\textit{Minimum Value}} 
& \textbf{\textit{Maximum Value}} 
& \textbf{\textit{Step}} \\
\hline
\textbf{$\alpha$} & 0.3 & 0.9 & 0.1 \\
\hline
\textbf{LSH Shingles} & 1-ngram & 5-gram & 1\\
\hline
\textbf{LSH Number of Permutations} & 16 ($2^4$) & 512 ($2^9$)  & $*2$ \\
\hline
\textbf{LSH Jaccard Threshold} & 0.3 & 0.9 & 0.1 \\
\hline
\textbf{$\beta$} & 0.3 & 0.9 & 0.1 \\
\hline

\hline
\end{tabular}
\egroup
\end{center}
\label{table:gs_range}
\end{table}

\begin{table*}[bt]
\caption{Optimal Parameter and Results Obtained by Grid Search Over Hyperparameters}
\begin{center}
\bgroup
\def\arraystretch{1.5}
\begin{tabular}{|c|>{\centering\arraybackslash}m{1.5cm}|>{\centering\arraybackslash}m{1.5cm}|>{\centering\arraybackslash}m{1.5cm}|>{\centering\arraybackslash}m{1.5cm}|>{\centering\arraybackslash}m{1.5cm}|>{\centering\arraybackslash}m{1.5cm}|>{\centering\arraybackslash}m{1.5cm}|}
\hline

\textbf{\textit{Dataset}}
& \textbf{\textit{$\alpha$}}
& \textbf{\textit{LSH Shingles}}
& \textbf{\textit{LSH Jaccard Threshold}}
& \textbf{\textit{LSH Number of Permutations}}
& \textbf{\textit{$\beta$}}
& \textbf{\textit{Number of Patterns}}
& \textbf{\textit{Loss}} \\
\hline
\textbf{Zookeeper} & 0.6 & 4 & 0.7 & 16 & 0.4 & 107 & 0.0050 \\
\hline
\textbf{HPC} & 0.7 & 4 & 0.7 & 32 & 0.4 & 309 & 0.0063 \\
\hline
\textbf{BGL} & 0.7 & 4 & 0.7 & 64 & 0.5 & 1470 & 0.0026 \\
\hline
\textbf{HDFS} & 0.6 & 4 & 0.7 & 16 & 0.4 & 42 & 0.0 \\
\hline

\hline
\end{tabular}
\egroup
\end{center}
\label{table:opti_sol}
\end{table*}

\section{Evaluating Privacy Preserving Filtering Framework}
\label{sec:privacy_preserving_eval}

\subsection{Experimental Setup}

In order to evaluate our privacy preserving framework, we generate a synthetic data set. The set of patterns used to generate the synthetic dataset are divided into two groups, one representing patterns from `success' and other representing patterns from `error'. The training set is generated using success patterns only, whereas the test set consists of both success and error patterns. We initialize our model with a fraction of success patterns and then learn rest of the success patterns in a privacy preserving manner. We calculate the false discovery rates and false negative rates for our model and compare them with those of a model trained on the complete training data to evaluate the correctness of our privacy preserving algorithm. 

\subsection{Dataset Generation}

First, we extract all the patterns from a set of log files obtained from different applications and designate 75\% of them as success patterns and remaining as error patterns. While generating the training set, we subdivide the success patterns into two sets. Patterns from the first set appear in every file of the training data, while the patterns from the other set are scattered among different files so as to ensure that no single file contains all the success patterns. Test set has similar distribution of success patterns along with randomly chosen mix of error patterns. Each of the training and test set contains eight files with fifteen-thousand lines each. The test and training data combined contain a total of 12968 patterns.

\subsection{Observations}

\subsubsection{Training with Privacy Preserving Model}

We initialized our clients by training them with one third of the training data. We sequentially process the remaining files in the training set to learn new patterns and update the set of pattern encodings. For the sake of this experiment, we skip the pattern selection part described in \ref{ssec:pp_server}. We observe that once the model learns encodings for a given file, it can identify the learned patterns with zero false positives and an average false negative rate of 0.14\%.

\subsubsection{Evaluation on Test Data}

Once trained with pattern encodings, we then run our model on the test set. Table \ref{table:pp_eval} shows a comparison of the performance of our privacy preserving model \textit{(M2)} against a model \textit{(M3)} trained on entirety of the training set and a model \textit{(M1)} trained only with one third train data set without our privacy preserving updates. We consider model \textit{(M3)} as the source of truth to calculate false negative and false discovery rate. As Table \ref{table:pp_eval} shows, privacy preserving learning helps in reducing the number of output patterns to half. We gain a twofold reduction in false discovery rate while the false negative rate stays at about two percent. Since, while debugging an application we do not want to miss any important line which may contain information relevant to problem identification, we tune the LSH parameters to minimize the false negative rate.

\begin{table}[htbp]
\caption{Evaluation of Privacy Preserving Learning Framework on Test Set of Our Sythetic Data}
\begin{center}
\bgroup
\def\arraystretch{1.2}
\begin{tabular}{|>{\centering\arraybackslash}m{2cm}|>{\centering\arraybackslash}m{1.5cm}|>{\centering\arraybackslash}m{1.5cm}|>{\centering\arraybackslash}m{1.5cm}|}
\hline
\textbf{\textit{Evaluation Metric}} & \textbf{\textit{M1}} &\textbf{\textit{M2}} & \textbf{\textit{M3}} \\
\hline
\textbf{Output Patterns} & 1588 & 774 & 563 \\
\hline
\textbf{False Negatives} & 0 & 12 & -\\
\hline
\textbf{False Negative Rate} & 0\% & 2.13\% & -\\
\hline
\textbf{False Positives} & 1023 & 222 & -\\
\hline
\textbf{False Discovery Rate} & 64.42\% & 28.68\% & - \\
\hline
\end{tabular}
\label{table:pp_eval}
\egroup
\end{center}
\end{table}


\section{Related Works}
\label{sec:related_works}

\subsection{Log Parsing}

Log parsing is a widely studied subject and various groups have attempted to devise several log parsing techniques in the past. Fu et al. \cite{fu} cluster log lines on the basis of weighted edit distance. On the other hand, LKE \cite{lke}, POP \cite{pop} and IPLoM \cite{iplom} attempt to segregate the log lines in various clusters such that all the lines in a cluster correspond to the same template. However, each of the above groups have used different techniques to cluster the log lines. LKE \cite{lke} use the edit distance between between each pair of log lines whereas POP \cite{pop} and IPLoM \cite {iplom} use number of tokens in a line to create initial clusters. POP \cite{pop} and IPLoM \cite {iplom} further split these initial clusters by identifying the tokens which occur most frequently at a given position and then splitting the clusters at these positions. IPLoM \cite{iplom} then find bijective relations between unique tokens, while POP \cite{pop} use relative frequency analysis for further subdivision of these clusters.

Vaarandi et al. \cite{sclt} use a different technique where they identify most commonly occurring tokens in the logs and use them to represent each template. Makanju et al. \cite{makanju} use a simple technique of iteratively partitioning logs into sub-partitions to extract log events. 

Logan \cite{logan}, Drain \cite{drain} and Spell \cite{spell} update log lines in order to reduce compute and memory overheads. Logan uses distributed compute with Apache Spark to reduce the training time. However, Zhu et al. \cite{zhu2018tools} show that these existing techniques deteriorate in both accuracy and performance when datasets contain large number of unique patterns. As we show in section \ref{sec:log_parsing_eval}, Delog scales notably well on datasets with large number of patterns.

\subsection{Log Processing for Problem Identification}

Use of machine learning and data mining techniques for anomaly detection in application logs is becoming increasingly common. Liang et al. \cite{liang} use a SVM classifier whereas Lou et al. \cite{lou} propose mining the invariants among log events to detect errors and anomalies.

Log3C \cite{log3c} uses a cascading clustering algorithm for clustering log sequences. These clustered sequences are then correlated with system Key Performance Indicators (KPI) using regression analysis. The clusters having high correlation with actual problems are identified. Similarly, Yuan et al. Beschastnikh et al.\cite{beschastnikh}, Shang et al. \cite{shang} and Ding et al. \cite{ding2012healing, ding2014mining} have proposed some other problem identification techniques but most relevant to our work is the log classification method proposed by Lin et al. \cite{lin}. They generate vector encoding of log sequences using Inverse Document Frequency (IDF) scores to calculate the similarity between client logs and stack traces of already known errors. However, this method requires log sequence level indicators of the application state and cannot be used in a privacy sensitive environment.




\section{Conclusion}
\label{sec:conclusion}

For most of the datasets, Delog fares almost two times better in training time performance as compared to the previous state of the art, Logan \cite{logan}. Moreover, the quality of patterns generated by Delog is also consistently better than the existing parsing algorithms. We use a minhash based LSH algorithm to obtain sub-linear complexity in number of patterns. The Thunderbird, Windows and Spark datasets used by us are the largest datasets to be used for log parsing so far. Delog is the only log parsing algorithm which is able to successfully parse the Thunderbird dataset in a respectable amount of time. We also perform exhaustive hyperparameter search across four datasets to demonstate the robustness of hyperparameter across datasets. 

Delog also uses a privacy preserving technique to constantly learn new patterns from client logs. User logs are homomorphically encrypted using a novel bloom filter based approach. We perform experiments on a synthetic dataset to demonstrate the efficacy of this approach in log filtering.


%

\bibliography{main}


%









\end{document}